\documentclass[journal]{IEEEtran}
\usepackage{graphics}
\usepackage{colortbl}
\usepackage{multicol}
\usepackage{diagbox}
\usepackage{lipsum}
\usepackage{color}
\usepackage[cmex10]{amsmath}
\usepackage{amsthm}
\usepackage{amssymb}
\usepackage{mathrsfs}
\usepackage{mathtools}
\usepackage{amsbsy}
\usepackage[colorlinks=true,bookmarks=false,citecolor=blue,urlcolor=blue]{hyperref}
\usepackage{xcolor}
\usepackage{setspace}
\usepackage{graphicx}
\usepackage{pstool}
\usepackage{epstopdf}
\usepackage{lettrine}
\usepackage{psfrag}
\usepackage{newclude}
\usepackage[normalem]{ulem}
\usepackage{latexsym}
\usepackage{algpseudocode}
\usepackage{algorithm,algpseudocode}
\usepackage{algorithmicx}
\usepackage{gensymb}
\usepackage{marginnote}
\usepackage{multirow}
\usepackage{bm}
\usepackage{wasysym}
\usepackage{cite,amsfonts}
\usepackage[T1]{fontenc}
\usepackage{comment}
\usepackage{tabularx}
\usepackage{caption}
\usepackage{csquotes}
\usepackage{float , subcaption}
\usepackage{makecell}
\usepackage{textcomp}
\usepackage{relsize}
\usepackage{hyperref}

\allowdisplaybreaks

\graphicspath{{figures/}}
\allowdisplaybreaks

\newcommand{\RNum}[1]{\uppercase\expandafter{\romannumeral #1\relax}}

\usepackage{mathtools}

\begin{document}
\title{An Accurate Model to Estimate 5G Propagation Path Loss for the Indoor Environment}
\author{\IEEEauthorblockN{Hassan Zakeri, Reza Sarraf Shirazi, and Gholamreza Moradi \thanks{Hassan Zakeri, Reza Sarraf Shirazi, and Gholamreza Moradi are with the Department of Electrical Engineering, Amirkabir University of Technology (Tehran Polytechnic), Tehran, Iran (E-mails: \{H.zakeri, Sarraf, Ghmoradi\}@aut.ac.ir).}
}}   
% make the title area
\maketitle
%%====> Abstract <===%%
\begin{abstract}
 This paper presents a new large-scale propagation path loss model to design a fifth-generation (5G) wireless communication system for indoor environments. Simulations for the indoor environment, for all polarization at non-line-of-sight (NLOS) and line-of-sight (LOS), which are performed per meter over a distance of 47 m between each of the separated transmitter antenna (TX) and the receiver antenna (RX) positions to compare better the proposed extensive flexible path loss model with previous models.

All the simulations are conducted at the Abu-rayhan buildings at the Amirkabir University of Technology. The results demonstrated that the simple presented model with a single parameter denoted ZMS can predict the expansive path loss over distance more accurately.
The values of the path loss exponent (PLE) for the LOS scenario are simulated and achieved at 3.63, 1.81, and 3.42 for the V-H, V-V, and V-Omni antenna polarizations, and for NLOS is 6.11, 4.21, and 5.23 at the 28 GHz frequency for all the polarization antenna type V-H, V-V, and V-Omni, appropriately. 
\end{abstract}
\begin{IEEEkeywords}
Fifth Generation, Indoor Propagation, Line-of-sight, Non-line-of-sight, Path loss models
\end{IEEEkeywords}
\IEEEpeerreviewmaketitle
\section{Introduction}
Fifth-generation (5G) technology is a wireless community branch that explores mobile radio systems to support new and advanced applications that require higher data rates, lower latency, higher reliability, and acceptable availability. 
It is important to note that a large number of connected devices are maintained simultaneously in this mobile generation.
Additionally, it is motivated by the tremendous growth of machine-to-machine (M2M) communications and their benefits and use in high technology structures in the Internet of Things (IoT) system.

In the last decade, different technologies are developed, such as the role of 5G in transportation in smart cities and in-vehicle communication \cite{guevara2020role}, creating the artificial system to control network traffic (CTN)  \cite{fu2018artificial}, permitting to create the protocol to increase the security of the application and software \cite{noohani2020review}, satellite communication \cite{wang2021location}, IoT \cite{knieps2022internet}, in biomedical to detect the tumors \cite{marriwala2022compact}, innovative sensors \cite{almasarani20215g}, Passive Optical Networks (PON) \cite{shah2021design}, energy harvesting \cite{sabban2022wearable}, Massive Multiple Input Multiple Output (MIMO) \cite{dicandia2021exploitation}, robotic control communication \cite{liu2021robotic}, and hybrid power systems for convenient cellular operation \cite{okundamiya2022optimization}. The "Big Three" of the coming wireless networks are small cell networks, mm-wave technology, and D2D communication \cite{imran2014challenges,kim2022analysis,nadeem2021efficient}.

Consequently, with the Big Three’s support, the devices using 5G networks will benefit from low propagation power, adequate bandwidth, and fast access delay. Therefore it converts to a high Signal to Interference Ratio (SIR) performance which is not precisely equivalent to today’s modern wireless system \cite{hindia2019enabling}.

Among various 5G enabling technologies, the mm-wave frequency range is the most efficient method to increase bandwidth \cite{singh2021potential}. 
This is due to the frequency band now in use, which is lower than 6 GHz that has not enough capacity to support high transmission rates. \cite{desai2021wideband}. 
The millimeter-wave spectrum at frequencies greater than 6 GHz has emerged as a possible applicant for the updated cellular 5G wireless communication \cite{tawa202128,ullah2021series,saha2021spectrum}. 
Therefore, the 5G cellular communication system’s millimeter-wave frequencies would reduce multimedia services. However, extensive classification, comprehensive characterization, and model construction in the mm-wave frequency range are required to build a generic and accurate model \cite{yang20185g}. 
%As a result, using millimeter-wave bands for 5G cellular communication systems would reduce multimedia services.

Recently, for both indoor and outdoor environments, the researchers presented different path loss models and algorithms for omni-directional and directional polarization based on measurements or artificial intelligence like machine learning such as Artificial Neural Networks (ANN), Support Vector Regression (SVR), and random forest can predict the amount of path loss propagation in different environments and situations.
 \cite{majed2018channel,genc2021new,hervis2022indoor}.

We present a novel method to calculate path loss in indoor environments accurately. All the simulations are carried out for the NLOS and LOS scenarios to compare better with the popular and new wide-range path-loss models. 
Samples of calculations from the simulated results were extracted using the CST Studio Site and then, the path loss model parameters were constructed using MATLAB to find the best results for the 5G wireless communication application.

The simulation is carried out for NLOS and LOS scenarios to compare popular and new wide-range path loss models. Also, it is performed for both NLOS and LOS scenarios to compare the expected and new extensive path loss models. After that, channel characterization and propagation modeling are investigated and examined based on the common and suggested path loss models for omni-directional and directional polarization.

The organization sequence of the sections in this study is as follows. The study’s methodology and the simulation setting are presented in Section II. Section III presents a wide-range characterization of single-frequency path loss models. Section IV presents the proposed models. The simulation analysis and results are presented and discussed in Section V.
Finally, in section VI, conclusions are provided.
\section{Simulation einvoirment}
\label{section: Simulation einvoirment}
As shown in Fig. 1, the simulation environment assumed concrete bricks and walls, tables, wooden shelves, drywall, glass, plastic, wooden doors, and indoor foliage, which makes it more representative. 

In accordance with ITU-R P.2040 regulations, the model also took into account the impact of the interconnection across building construction materials and operational frequencies, as indicated in Table 1.

\begin{table}[ht]
\centering
\caption{Material properties at 24-39 GHz  \cite{lee2017permittivity}.}
\begin{tabular}[t]{lcc}
\hline
Material  & $\epsilon_r$ & $\sigma$ \\
\hline
Concrete  & 5.31 & 0.48  \\
Glass     & 6.27 & 0.23  \\
Wood      & 1.99 & 0.17  \\
Drywall   & 2.94 & 0.12  \\
\hline
\end{tabular}
\label{8th}
\end{table}%
Simulations took place in a simulated environment on the 8th floor of the Abu-rayhan buildings at the Amirkabir University of Technology.

\begin{figure}[!ht]
	\centering\includegraphics[scale=0.22]{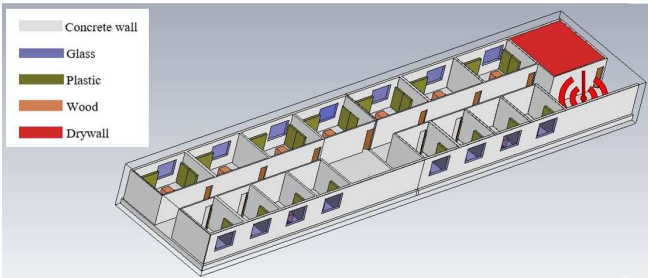}
	\caption{3D view of the simulated environment for the 8th floor Abu-rayhan buildings with the position of the transmitter antenna at the Amirkabir University of Technology. }
	\label{8thhh}
\end{figure}

Vertical polarization (V-V) is used to evaluate co-polarization in directional path loss models using horn antennas as transmitters and receivers. The TX antenna is polarized in a vertical state for the cross-polarization assessments, while the RX antenna is polarized in horizontal mode (V-H). The TX horn antenna and RX Omni-antenna are also polarized in vertical mode for the omni-directional path loss model. TX height is determined to be 2.5 m above the ground, considered an indoor hotspot on the wall. It is important to note that the ceiling height is 3 m, and the RX antenna’s height is fixed at 1.5 m (typical phone height handset).
 
The building’s three TX sites and 47 RX sites were selected for power quantification. Additionally, all RX sites are dispersed between NLOS and LOS, with TX-RX disconnection ranging between 1.9 m and 45.7 m. It is important and it is necessary to indicate that the ground floor area measures approximately 51 m × 12 m. As a result, the transmitter and receiver were disconnected by 1 m at zero points (i.e., the first point of the assessment). The TX antenna is placed in an immovable position in the ground floor hallway of the building for the assessment procedure.

The evaluation started with the RX 1.09 m from the transmitter, in which the received signal was recorded and the RX was stationary. The calculated samples by considering reflection, diffraction, and scattering from the materials, were then extracted.

Then, the parameters of the path loss models were constructed by employing large-scale methods to obtain the best match for the samples because path loss models are needed to determine the signal strength along a route. 
The Root Mean Square Error (RMSE) between the CST Studio Site, simulated data, and generated path loss model data were used as performance indicators. It is important to note that the better the model, the lower the RMSE.

%%%%%%%%%%%%%%%%%%%%%%%%%%%%%%%%%%%%%%%%%%%%%%%%%%%%%%%%%%%%

\section{LARGE-SCALE CHARACTERIZATION}
\label{section:LARGE-SCALE CHARACTERIZATIONr}

The close-in (CI) free-space path loss model is a typical and useful path loss model. This model is frequently used to display how a channel evaluated by the environment might affect the system. 
The fading behavior, also known as the power attenuation parameter, was simultaneously evaluated using the path loss model as a distance and frequency function.

The CI free-space path loss model has the following notation.
 \cite{sun2016investigation}: 

\begin{equation}
PL^{CI}(f,d)[dB]= FSLP(f,d_0)+ 10n\log(\frac{d}{d_0})+X_{\sigma}^{CI}
\end{equation}
 
 and the $FSLP(f,d_0)$ is
 
 \begin{equation}
 FSLP(f,d_0)= 20\log (\frac{4\pi d_0}{\lambda})
\end{equation}

where $n$ stands for the $PLE$, $X_{\sigma}^{CI}$ represents a Gaussian random variable,  $\lambda$ is the carrier wavelength that depends on the value of the frequency, $\sigma$ illustrates the standard deviation in dB, and $d_0 = 1m$ specifies the physical reference distance.

The appendix shows CI's methodology to minimize shadow fading (SF).

In order to get the optimum minimum error fit of the combined path losses, the floating-intercept (FI) path loss model that is used in 3GPP and WINNER II standards depends on the $\alpha$ and the $\beta$. The FI is denoted as \cite{maccartney2015indoor}:

\begin{equation}
PL^{FI}(f,d)[dB]= \alpha + 10\beta \log(d)+X_{\sigma}^{FI}
\end{equation}

where $\alpha$ represents the floating intercept in dB, $\beta$ indicates the line slope, and $X_{\sigma}^{FI}$ indicates the extensive signal variability received versus distance along the direct path.

%%%%%%%%%%%%%%%%%%%%%%%%%%%%%%%%%%%%%%%%%%%%%%%%5
%%%%%%%%%%%%%%%%%%%%%%%%%%%%%%%%%%%%%%%%%%%%%%%%%%%

\section{PROPOSED MODEL FOR INDOOR 5G
COMMUNICATION}
\label{section: PROPOSED MODEL FOR INDOOR 5G
COMMUNICATION}

The progressive narrow-beam mm-wave technology can be used when receivers combine most energy from different angles of arrival.

The currently proposed approach has made the system design and performance evaluations simpler. In high-density environments with numerous obstructions, the conventional single-slope path loss function could not reflect the path loss reliance on separateness distance. The path loss value is divided into NLOS and LOS coefficient factors and is correlated to distance $d$. 

The presented model evaluations give practically  based and accurately calculated path loss data points for the reference distance $d_0 = 1m$, much like the CI path loss model.
As shown in (5), (6), and (7), respectively, the two corrections of the coefficient are recommended for both the NLOS and LOS environments for the mm-wave bands. 

The simulated data determined parameter ZMS anticipated offering a good match path loss.

The study suggested the following general propagation model based on stochastic distribution functions for NLOS and LOS scenarios, taking into consideration the function of TX-RX disconnection distance and path loss valuation modification: 

\begin{equation}
\begin{split}
PL^{ZMS}(f,d)[dB]= FSLP(f,d_0)+ \\ 10n\log(\frac{d}{d_0})+
ZMS(f,d) +X_{\sigma}^{ZMS}
\end{split}
\end{equation}

For the ZMS(f,d) condition, we have (see the appendix for how to minimize the SF for the proposed model):

ZMS(f,d) in V-V
\begin{small}
\begin{equation}
%\begin{align}
\begin{split}
LOS : &  0 \\ 
NLOS :& \frac{\sum_{n=t+1} ^{l} (\sqrt{ P(f,d_t)_{V-V} \times d_t } - \sqrt{ P(f,d_r)_{V-Omni \times d_t }})^2}{D_L}
\end{split}
%\end{align*}
\end{equation}
\end{small}
ZMS(f,d) in V-H
\begin{small}
\begin{equation}
%\begin{align*}
\begin{split}
LOS : & \frac{\sum_{n=0} ^{t} (\sqrt{ P(f,d_t)_{V-H} \times d_t } - \sqrt{ P(f,d_r)_{V-Omni} \times d_r })^2}{D_L} \\ 
NLOS :& \frac{\sum_{n=t+1} ^{l} (\sqrt{ P(f,d_t)_{V-H} \times d_t } - \sqrt{ P(f,d_r)_{V-V} \times d_r })^2}{D_L}
\end{split}
%\end{align*}
\end{equation}
\end{small}
ZMS(f,d) in V-Omni
\begin{small}
\begin{equation}
%\begin{align*}
\begin{split}
LOS : &  \frac{\sum_{n=0} ^{t} (\sqrt{P(f,d_t)_{V-Omni} \times d_t } - \sqrt{P(f,d_r)_{V-V} \times d_r })^2}{D_L} \\ 
NLOS : & \frac{\sum_{n=t+1} ^{l} (\sqrt{ P(f,d_t)_{V-H} \times d_t } - R_{NL})^2}{D_L} 
\end{split}
%\end{align*}
\end{equation}
\end{small}
Where $R_{NL}$ is an adjust factor

\begin{equation}
R_{NL}=\left|\frac{(\sqrt{ P(f,d_t)_{V-H} \times d_t } - \sqrt{ P(f,d_r)_{V-V} \times d_r })}{2}\right|
\end{equation}

ZMS model represents the combined simulated points at identical separation distances for the LOS and NLOS scenarios; $PL^{ZMS}(f,d)$ denote the path loss at $d_0 = 1m$,
$t$ indicates the distance index, and $l$ indicates the longest separation distance.
The $D_L$ is the longest distance between the TX and the RX antenna.
It is important to note that the $d_t$ is the distance from the TX and $d_r$ is the average distance from other RX antennas. 
\begin{figure}
\centering
\begin{subfigure}{\linewidth}
\label{fig:figure:im1} %% label for first subfigure
\centering\includegraphics[width=\linewidth]{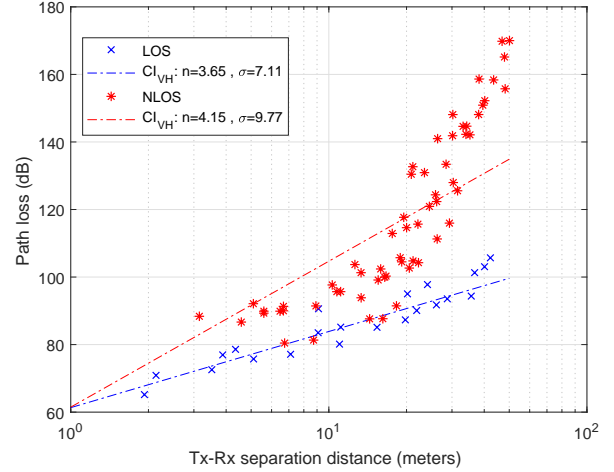}
\subcaption{CI\_vh}
\end{subfigure}
\\[5mm]
\begin{subfigure}{\linewidth}
\label{fig:figure9:im22} %% label for second subfigure
\centering\includegraphics[width=\linewidth]{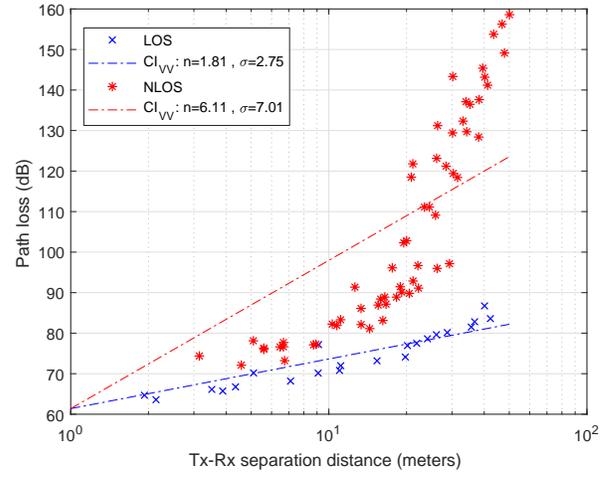}
\subcaption{CI\_vv}
\end{subfigure}
\\[5mm]
\begin{subfigure}{\linewidth}\label{fig:figure10:im222} %% label for second subfigure
\centering\includegraphics[width=\linewidth]{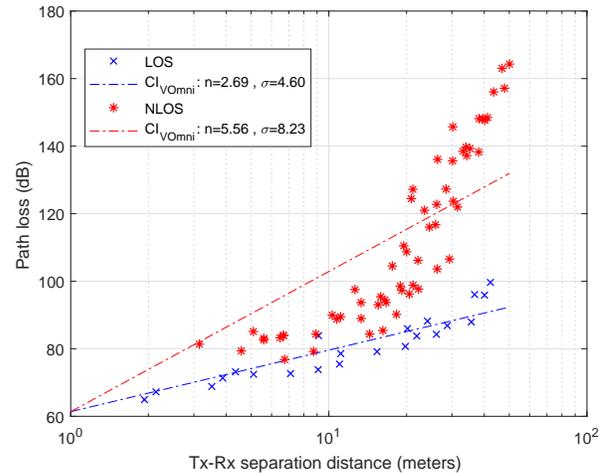}
\subcaption{CI\_omni}
\end{subfigure}
\\[5mm]
\caption{CI path loss model for V-H, V-V, and V-Omni for LOS and NLOS at 28 GHz}
\label{fig:figure11} %% label for entire figure
\end{figure}

% =================================================================================
\section{results and discussion}
\label{section:results}
\subsection{Single-Frequency Path Loss Models}
 
Fig. 2 demonstrates the omni-directional and directional path loss for the CI model at 28 GHz in the NLOS and LOS environment for all antenna polarizations: V-H, V-V, and V-Omni, respectively.

For V-H, V-V, and V-Omni antenna polarization, the values of the PLE in the LOS case study are 3.63, 1.81, and 3.42 at a frequency of 28 GHz. By mathematical calculation of LOS, PLE is 1.81 at 28 GHz, as shown in V-V antenna polarization, which is less than the conceptual free-space PLE of 2, demonstrating that the ground and ceiling bounce reflections cause constructive interference with an indoor mm-wave propagation channel. Also, the waveguide effects in halls and corridors have a LOS directional PLE that is not dependent on the amount of frequency. It is essential to note that the PLE is higher for the V-H antenna polarization in LOS than in V-V.
There is a noticeable depolarization effect at these frequencies in both LOS and NLOS inside situations, as shown by the  28 GHz PLE values of 5.29 and 4.54 for V-V and V-H antenna polarization, respectively.

Table II lists the parameters of the CI path loss model for NLOS and LOS at 28 GHz. For the 28 GHz frequency band, the shadow factor of the mean path loss line for the LOS V-V arrangement is 2.75 dB.

\begin{table} [h!]
    \centering
    \caption{CI path loss model parameters }
    \begin{tabular}{ lcccc }
    \hline
     & \multicolumn{2}{c}{LOS} & \multicolumn{2}{c}{NLOS}
     \\
     Polarization & $n$ & $\sigma$ & $n$ & $\sigma$ \\
    \hline 
V-V    & 1.81 & 2.75 & 5.29 & 7.56 \\
V-H    & 3.59 & 7.25 & 4.54 & 9.25 \\
V-Omni & 2.69 & 4.60 & 5.56 & 8.23 \\
    \hline
    \end{tabular}
    
    \label {table:1}
    \end{table}

The V-Omni and V-V configuration standard deviations for the NLOS and LOS simulations varied from 2.7 dB to 9.2 dB between the examined frequencies, demonstrating that the strongest received signal fluctuates more than the average received signal power across all TX-RX separate distances.

The FI path loss model results for the 28 GHz frequency bands in NLOS and LOS environments are demonstrated in fig. 3 for the V-H, V-V, and V-Omni antenna polarizations, respectively.

\begin{figure}
\centering
\begin{subfigure}{\linewidth}\label{fig:figure1:im11} %% label for first subfigure
\centering\includegraphics[width=\linewidth]{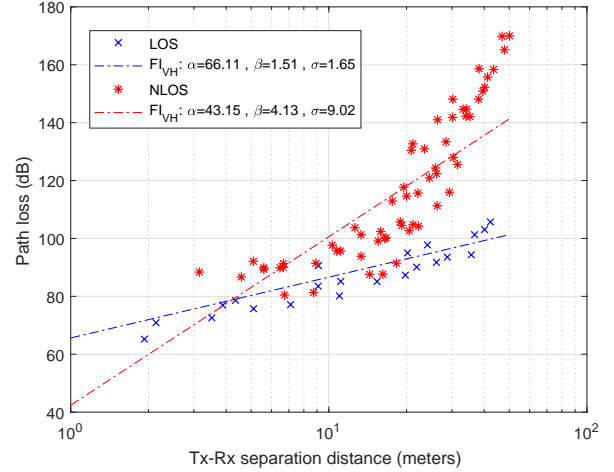}
\subcaption{FI\_vh}
\end{subfigure}\\[5mm]
\begin{subfigure}{\linewidth}\label{fig:figure2:2222} %% label for second subfigure
\centering\includegraphics[width=\linewidth]{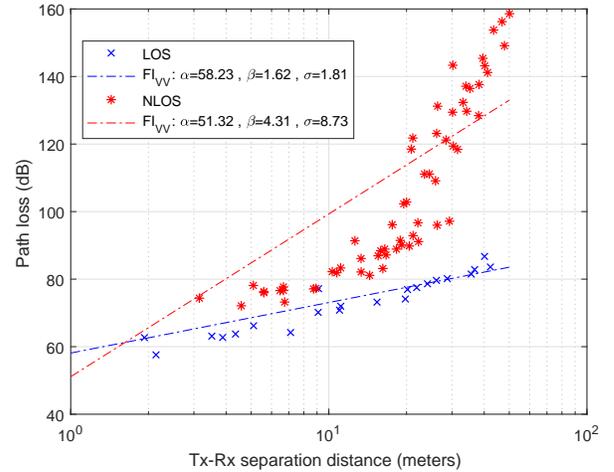}
\subcaption{FI\_vv}
\end{subfigure}\\[5mm]
\begin{subfigure}{\linewidth}\label{fig:figure3:im2333} %% label for second subfigure
\centering\includegraphics[width=\linewidth]{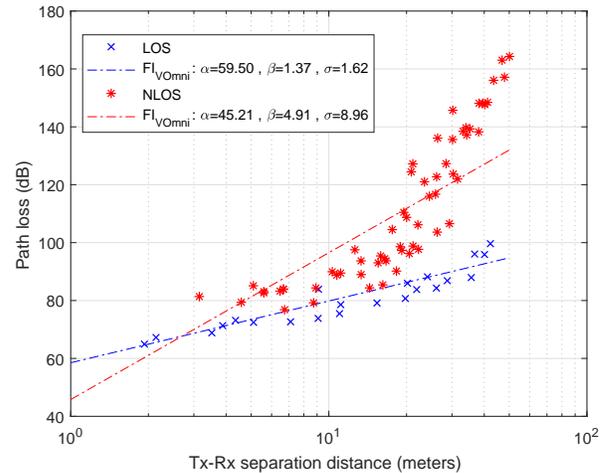}
\subcaption{FI\_omni}
\end{subfigure}
\\[5mm]
\caption{FI path loss model for V-H, V-V, and V-Omni for LOS and NLOS
at 28 GHz.}
\label{fig:figure4} %% label for entire figure
\end{figure}

At 28 GHz, it was 58.23 dB compared to the 61.41 dB theoretical FSPL at 1 m. In LOS V-V, the $\alpha$ values can vary from free-space path losses.
Table III illustrates the parameters for the FI path loss model.

\begin{table} [h!]
    \centering
    \caption{FI path loss model parameters }
    \begin{tabular}{ lcccccc }
    \hline
     & \multicolumn{3}{c}{LOS} & \multicolumn{3}{c}{NLOS}
     \\
     Polarization & $\alpha$ & $\beta$ & $\sigma$ & $\alpha$ & $\beta$ & $\sigma$  \\
    \hline 
V-V    & 58.23 & 1.62 & 1.81 & 51.32 & 4.31 & 8.73 \\
V-H    & 66.11 & 1.51 & 1.65 & 43.15 & 4.13 & 9.02 \\
V-Omni & 59.50 & 1.37 & 1.62 & 45.21 & 4.91 & 8.96 \\
\hline
    \end{tabular}
    
    \label {table:2}
    \end{table}
    
The floating-intercept values in the NLOS environment ranged from 45.21 dB to 51.32 dB and are not frequency-dependent. The $\beta$ of the mean least-square fit line for the 28 GHz band with cross- and co-polarized antennas is near to the free space ($\beta$ = 2) does not necessarily mean that the NLOS signals offer significantly more power reduction with distance respected the free-space signals.

The extremely sensitive FI model is demonstrated by the incredibly low LOS $\beta$ of 1.62 at 28 GHz V-V and 1.51 for V-H antenna polarization, where PLE values of both polarizations suggest a negligible increase in path loss as distance rises.

\subsection{Proposed Model Analysis}

The simulation results are consistent with the model shown in fig. 4 for the V-H, V-V, and V-Omni antenna polarizations in NLOS and LOS scenarios for the 28 GHz band.

\begin{figure}
\centering
\begin{subfigure}{\linewidth}\label{fig:figure5:im1111} %% label for first subfigure
\centering\includegraphics[width=\linewidth]{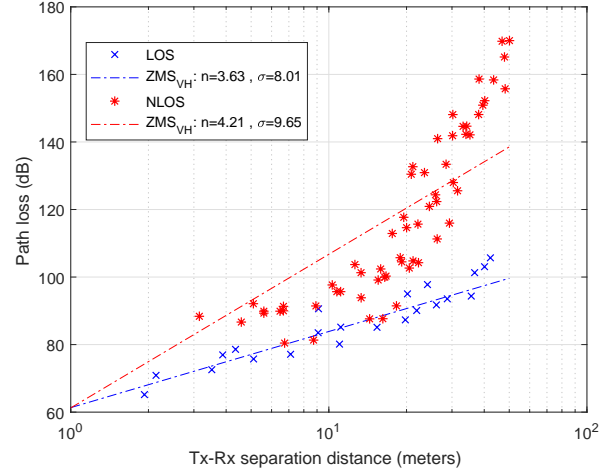}
\subcaption{ZMS\_vh}
\end{subfigure}\\[5mm]
\begin{subfigure}{\linewidth}\label{fig:figure6:im2654} %% label for second subfigure
\centering\includegraphics[width=\linewidth]{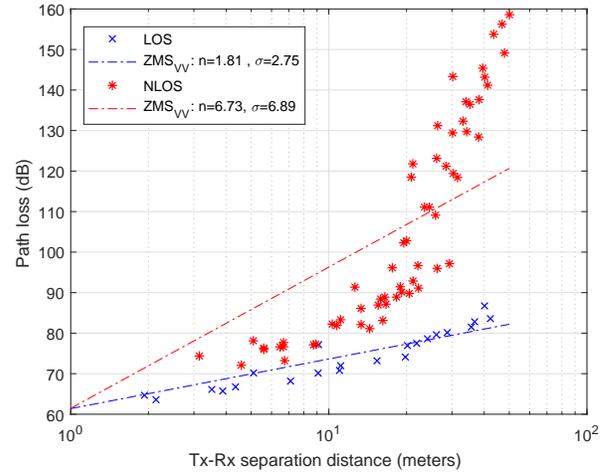}
\subcaption{ZMS\_vv}
\end{subfigure}\\[5mm]
\begin{subfigure}{\linewidth}\label{fig:figure7:im2} %% label for second subfigure
\centering\includegraphics[width=\linewidth]{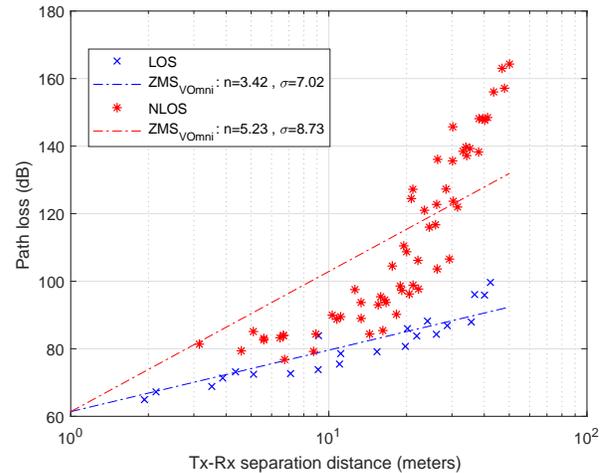}
\subcaption{ZMS\_omni}
\end{subfigure}
\\[5mm]
\caption{ZMS path loss model for V-H, V-V, and V-Omni for LOS and NLOS
at 28 GHz.}
\label{fig:figure8} %% label for entire figure
\end{figure}

The correction factor for any environment is the only parameter that must be estimated. 
Table IV demonstrates that the simulation results indicate considerable variation in the path losses for various antenna polarizations for both scenarios in NLOS and LOS in the frequency band.

\begin{table} [h!]
    \centering
    \caption{ZMS path loss model parameters }
    \begin{tabular}{ lcccc }
    \hline
     & \multicolumn{2}{c}{LOS} & \multicolumn{2}{c}{NLOS}
     \\
     Polarization & $n$ & $\sigma$ & $n$ & $\sigma$ \\
    \hline 
V-V    & 1.81 & 2.75 & 6.73 & 6.89 \\
V-H    & 3.63 & 8.01 & 4.21 & 9.65 \\ 
V-Omni & 3.42 & 7.02 & 5.23 & 8.73 \\
    \hline
    \end{tabular}
    
    \label {table:4}
    \end{table}

The proposed model for path loss in the LOS situation for 5G wireless communication exhibits the same phenomenon.
As a result, the suggested model for the 28 GHz band is accurate and simple. In LOS at 28 GHz, the PLE values for the offered model are 3.63, 1.81, and 3.42 for V-H, V-V, and V-Omni antenna polarizations, respectively, compared to the CI path loss model, compared to PLEs of 3.59, 1.81, and 2.69 at the same antenna polarization for the CI path loss model. 
The introduced model’s PLE values for the V-H, V-V, and V-Omni antenna polarizations at NLOS 28 GHz are 6.11, 4.21, and 5.23, respectively, compared to the CI path loss model’s PLEs of 5.29, 4.54, and 5.56 for the same antenna polarization at the same location.

Additionally, the standard deviation ($\sigma$) in the majority of presented models is lower than the average deviation values for the CI path loss model.
The suggested model improves the CI path loss model in terms of path loss prediction for $\sigma$ and $n$ parameters for 28 GHz in NLOS and LOS scenarios. 

Table V demonstrates the comparison of the offered and CI model parameters at the 28 GHz frequency
band.

\begin{table} [h!]
    \centering
    \caption{Comparison between ZMS and CI path loss model parameters.}
    \begin{tabular}{ lccccc }
    \hline
    & & \multicolumn{2}{c}{LOS} & \multicolumn{2}{c}{NLOS}
     \\
  Model   & Polarization & $n$ & $\sigma$ & $n$ & $\sigma$ \\
  \hline
    & V-V    & 1.81 & 2.75 & 6.73 & 6.89 \\
ZMS & V-H    & 3.63 & 8.01 & 4.21 & 9.65 \\ 
    & V-Omni & 3.42 & 7.02 & 5.23 & 8.73 \\
    \hline
   & V-V    & 1.81 & 2.75 & 5.29 & 7.56 \\
CI & V-H    & 3.59 & 7.25 & 4.54 & 9.25 \\
   & V-Omni & 2.69 & 4.60 & 5.56 & 8.23 \\
    \hline

    \end{tabular}
    
    \label {table:3}
    \end{table}

\section{Conclusion}

This study presents a novel path loss model for the 28 GHz band in an indoor environment. The impacts of path loss when The 5G signals were sent over the frequency band and studied using this simulation exercise. The CI model simulated the data more effectively than the 3GPP models for the single-frequency path-loss model. 
The results of the FI model reveal that the path loss model is not sufficiently precise to account for the physical consequences of environmental loss across a distance. Therefore, the FI model does not precisely capture the path-loss performance channel in a LOS or NLOS environment. 
Additionally, the small difference in the standard deviation illustrates that the FI model cannot be as well matched for closed-form analysis as a simpler physical-based CI model. 
In addition, NLOS and LOS indoor environments are developed using the suggested model with a newly calculated parameter based on actual field environment power simulation experiments.

Furthermore, when analyzed and evaluated by comparing the CI and FI models, the suggested model showed an average increment compared with the other models discussed. 
This demonstrates that a reliable communication link can be defined on a 28 GHz band for indoor use. 
Moreover, it can reasonably estimate the mm-wave path loss as a function of the distance, frequency, and other system-specific environmental parameters.

\appendix

This appendix provides mathematical derivations for the CI, FI, and the proposed model ZMS closed-form solutions by finding model parameters to minimize the SF standard deviation. 

\subsection{The CI path loss model}
By considering the reference distance equivalent to 1 m, the CI model’s expression is calculated by \cite{sun2016investigation}:

\begin{equation}
PL^{CI}(f,d)[dB]= FSLP(f,d_0)+ 10n\log(\frac{d}{d_0})+X_{\sigma}^{CI}
\end{equation}

So, the SF is achieved by:
\begin{equation}
X_{\sigma}^{CI} = P-nL 
\end{equation}

where the $L$ and $P$ indicates the $10n\log(d)$ and the $PL^{CI}(f,d)[dB]- FSLP(f,d_0)$, respectively.

Therefore the SF deviation is:

\begin{equation}
\sigma^{CI}=\sqrt((\sum (X_{\sigma}^{CI})^2/N )= \sqrt(\sum(P-nL)^2/N )
\end{equation}

for minimizing the SF the derivation of the numerator based on $n$ should be minimized:  

\begin{equation}
\frac{d\sum(P-nL)^2}{dn}=\sum 2L(nL-P)=0
\end{equation}
 
 So the result is:
 
\begin{equation}
n=\frac{\sum PL}{\sum L^2}
\end{equation}

\subsection{The FI path loss model}

For the FI model \cite{maccartney2015indoor}:

\begin{equation}
PL^{FI}(f,d)[dB]= \alpha + 10\beta \log(d)+X_{\sigma}^{FI}
\end{equation}

Thus the SF is:

\begin{equation}
X_{\sigma}^{FI} = C-\beta L-\alpha
\end{equation}

where the $L$ and $C$ indicates the $10n\log(d)$ and the $PL^{CI}(f,d)[dB]$, respectively.

Therefore the SF deviation is:

\begin{equation}
\sigma^{FI}=\sqrt((\sum (X_{\sigma}^{FI})^2/N )= \sqrt(\sum(C-\beta L-\alpha)^2/N )
\end{equation}

the term $(C-\beta L-\alpha)^2$ should be minimized to find the best adaptive SF:

\begin{equation}
\frac{d\sum(C-\beta L-\alpha)^2}{d\alpha}=\sum 2(\beta L-C+\alpha)=0
\end{equation}

and 

\begin{equation}
\frac{d\sum(C-\beta L-\alpha)^2}{d\beta}=\sum 2L(\beta L-C+\alpha)=0
\end{equation}

Finally, the $\alpha$ and the $\beta$ are:

\begin{equation}
\alpha=\frac{\sum L \sum LC - \sum L^2 \sum C}{(\sum L)^2 -N\sum L^2}
\end{equation}

and

\begin{equation}
\beta=\frac{\sum L \sum C - N \sum LC}{(\sum L)^2 -N\sum L^2}
\end{equation}

\subsection{The ZMS path loss model}

For the proposed ZMS model with a reference distance of 1 m,

\begin{equation}
\begin{split}
PL^{ZMS}(f,d)[dB]= FSLP(f,d_0)+ \\ 10n\log(\frac{d}{d_0})+
ZMS(f,d) +X_{\sigma}^{ZMS}
\end{split}
\end{equation}

So, the SF is achieved by:
\begin{equation}
X_{\sigma}^{ZMS} = P-nL -ZMS
\end{equation}

and the SF deviation is:

\begin{equation}
\sigma^{CI}=\sqrt((\sum (X_{\sigma}^{ZMS})^2/N )= \sqrt(\sum(P-nL-ZMS)^2/N )
\end{equation}

for minimizing the SF the derivation of the numerator based on $n$ should be minimized:  

\begin{equation}
\frac{d\sum(P-nL-ZMS)^2}{dn}=\sum 2L(nL-P+ZMS)=0
\end{equation}
 
 So the final result for the proposed model is:
 
\begin{equation}
n=\frac{\sum PL-\sum((ZMS)L)}{\sum L^2}
\end{equation}

%%====> References <===%%
\bibliographystyle{IEEEtran}
\bibliography{5g.bib}
\end{document}